# Stacking-configuration-preserved Graphene Quantum Dots Electrochemically Obtained from CVD Graphene

Santiago D. Barrionuevo,[a] Federico Fioravanti,[b] Jorge M. Nuñez,[c-g] David Muñeton Arboleda,[a] Gabriela I. Lacconi,[b] Martin G. Bellino,[i,*] Myriam H. Aguirre,[fgh] and Francisco J. Ibañez.[a,*]

[a] Instituto de Investigaciones Fisicoquímicas, Teóricas y Aplicadas (INIFTA). Universidad Nacional de La Plata - CONICET. Sucursal 4 Casilla de Correo 16 (1900) La Plata, Argentina.

[b] INFIQC-CONICET, Dpto. de Fisicoquímica, Facultad de Ciencias Químicas, Universidad Nacional de Córdoba, Ciudad Universitaria, (5000) Córdoba, Argentina.

[c] Resonancias Magnéticas-Centro Atómico Bariloche (CNEA, CONICET) S. C. Bariloche 8400, Río Negro, Argentina.

[d] Instituto de Nanociencia y Nanotecnología, CNEA, CONICET, S. C. Bariloche 8400, Río Negro, Argentina.

[e] Instituto Balseiro (UNCuyo, CNEA), Av. Bustillo 9500, S.C. de Bariloche 8400, Río Negro, Argentina.

[f] Instituto de Nanociencia y Materiales de Aragón, CSIC-Universidad de Zaragoza, C/ Pedro Cerbuna 12, 50009, Zaragoza, Spain.

[g] Laboratorio de Microscopías Avanzadas, Universidad de Zaragoza, Mariano Esquillor s/n, 50018, Zaragoza, Spain.

[h] Dpto. de Física de la Materia Condensada, Universidad de Zaragoza, C/ Pedro Cerbuna 12, 50009, Zaragoza, Spain.

[i] Instituto de Nanociencia y Nanotecnología CNEA-CONICET, Av. Gral. Paz, 1499, San Martín, Buenos Aires, B1650, Argentina

* Author to whom correspondence should be addressed:

Dr. Martin G. Bellino

mbellino@conea.gov.ar

Dr. Francisco J. Ibañez

fjiban@inifta.unlp.edu.ar

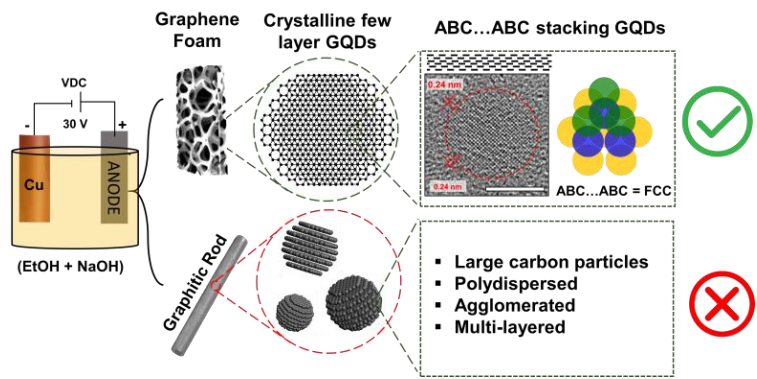

**TOC.** The scheme shows that as-synthesized GQDs conserved the crystalline and few-layered stacking structure from CVD graphene grown on Ni foam. When using a graphitic rod, agglomerated and polydisperse structures are observed.

**ABSTRACT.** *The layer stacking morphology in nanocarbons is paramount for achieving new properties and outperforming applications. Here, we demonstrate that graphene quantum dots (GQDs) retained crystallinity and stacking structure from CVD graphene grown on Ni foam. Our results reveal that GQDs subdomains comprise a few layer graphene structure in AB…AB and ABC…ABC stacking configuration. HR-TEM images along with a multiple-approach characterization (XRD, XPS, UV-vis, AFM, and ATR-IR) exhibit ~3.0 to ~8.0 nm crystalline graphene quantum dots (GQDs) with 2-6 graphene layers in disk-shape structure. The UV-vis profiles show changes in color of the dispersion (from colorless to red) during and after the electrochemistry suggesting a systematic electrooxidation of graphene into smaller, highly crystalline, and more complex $sp^2/sp^3$ structures. Importantly, a control experiment performed under the same conditions but with a graphitic rod exhibited large, polydisperse, and multi-layer carbon structures. This work demonstrates a relatively easy electrochemical synthesis to obtain GQDs retaining the pristine and in turn, distinctive structure of graphene grown on Ni foam.*

# INTRODUCTION.

Graphene quantum dots (GQDs) have attracted considerable interest as a fascinating and innovative zero-dimensional material in the carbon family, boasting exceptional optical, chemical, physical, electrical, and biological properties.[1] Small carbon structures measuring less than 10 nm, hold great promise for optoelectronic and electrochemical applications.[2–8] Their tunable properties, achieved through size and functional groups, make them great candidates for applications in ions detection,[9,10] bio-imaging,[11] photocatalysis,[7,12] high performance supercapacitors,[13] and the design of efficient photoanodes.[14] There is a large number of bottom-up and top-down methods to obtain GQDs however, most of them have poor control over the product of synthesis. For bottom-up approaches, the synthesis using polycyclic aromatic hydrocarbons (PAHs) is, perhaps, the one with the highest control over the structure and crystallinity of the final product but it requires long chemical synthesis steps.[15] Top down electrochemical approaches seem quite promising because multiple parameters such as solvent, pH, voltage, and more importantly; the composition/structure of the anode can be altered in order to improve control over the product. The selection of the carbon precursors usually includes bulk materials such as coke, graphite, HOPG, graphitic rods, etc.[16–18] Another promising alternative, although unusual, relays on using graphene-based precursors since the obtained product is expected to at least resemble some of the original structure. Just few groups obtained GQDs from already formed carbon nanostructures including 1D carbon nanotubes,[19] 2D,[20] and 3D[14,21] graphene foams. For instance, Shinde and Pillai demonstrated control over the size by two steps oxidation/reduction of carbon nanotubes which led to 3-8 and 23 nm diam. graphene dots.[19] Later, Chen and co-workers obtained ~3.0 nm diam. crystalline structures from the electrochemical exfoliation of 3D graphene in the presence of organic solvents and ionic liquids.[21] Our group followed a similar approach to Chen and co-coworkers although the synthesis was completely performed in polar solvents to ultimately be applied in energy-related uses[14] and selective ion detection in real water samples[9] and wine berberages.[10]

The general top-down mechanism consists in promoting the oxidation of C=C structures by breaking the sp$^2$-hybrization at defects sites within the basal plane of the carbon framework.[22] Nevertheless, when the electrochemistry is performed in the presence of a second source of carbon (i.e.; small molecules such as ethanol) a bottom-up mechanism consisting in the electrooxidation of alcohols takes place as well.[9,14,23,24] That was one of the reasons why we recently explored the electrooxidation of ethanol on the surface of a bare Ni foam which led to ultra-small (3.0 nm diam.) and crystalline carbon nanostructures.[12] The possibility of two mechanism occurring simultaneously have been somehow underestimated.[19,25–27] Moreover, the inner structure of the starting carbon material is of crucial importance because it determines the optoelectronic properties of the final product. In line with that, the number of layers, rotation between them, and the stacking order has proved to be relevant toward the electronic transport and optical properties of useful nanocarbons. Recently, twisted bilayer[28] and ABC rhombohedral[29,30] graphene have shown superconductivity and strange correlation phenomena opening the possibility of new nanoelectronics research. In addition, chemical vapor deposition (CVD) has enabled the successful large-scale growth of high-quality stacking superstructures of graphene on metals.[31,32] Very few groups have demonstrated in-surface fabrication of GQDs from high quality CVD graphene layers.[33,34] For instance, Kang el al. fabricated 12, 14, 16, and 27 nm diam. GQDs by assembling Au nanoparticles on top of a graphene film (used as a mask) followed by etching the graphene to ultimately dissolve the Au to leave behind small graphene domains (i.e.; GQDs) which preserved the diameter of Au phantom.[34] A great challenge would be to fabricate colloidal GQDs capable of retaining the stacking structures of the CVD graphene precursor with smaller domains and new functionalities in a simple and reproducible manner.

In this work we obtain GQDs that conserved the high quality and crystalline stacking order structure after the electrooxidation of CVD-graphene-covered 3D Ni-foams used as ultrathin carbon precursor. During the synthesis there is a systematic oxidation of the graphene foam into smaller $sp^2$ subdomains that keeps evolving even after the synthesis is completed, as indicated by the distinct changes in color from yellow to red. We demonstrate that these small subdomains (i.e.; GQDs) retained crystallinity, structure, and overall quality from the CVD graphene precursor. Results exhibit a main population of ~8.0 nm diam. crystalline GQDs with 2-6 graphene layers in a disk-shape structure. Our results reveal that GQDs subdomains retained the few layer configuration exhibiting AB…AB as well as ABC…ABC stacked graphene layers. Moreover, we performed a control experiment replacing the CVD graphene by a graphite rod and observed amorphous, bulk, and polydisperse carbon structures. This clearly illustrates the capability of "oxidation scissor" to preserve structural order features of CVD graphene layers to attain sophisticated high quality carbon nanostructures. We also explored the influence of the solvent used during the electrosynthesis which critically impacted on the solubility of the GQDs into aqueous media increasingly important for medical and sensors applications. From these findings, it is expected that electrochemistry will promote even the practical synthesis of exotic heteroatom 2D quantum dots from van der Waals heterostructures.

**EXPERIMENTAL METHODS.**

**Reagents**. All chemical reagents were used as received. Ethanol (99%), sodium hydroxide, were purchased in Sigma-Aldrich Argentina. Aqueous solutions were prepared with ultrapure water (18.2 mΩ resistivity). Nickel foam sheets with 1.6 mm thickness and 87 % porosity was purchased from MTI Corp. (Richmond, CA, USA). $H_2$ (99.999 %) and $CH_4$ (99.999 %) were purchased locally from Linde, Argentina.

**Synthesis of Graphene Quantum Dots.** GQDs were synthesized through electrochemical cleavage of graphene already grown on Ni foam[36] following the same protocol recently published by our group.[14] Briefly a CVD graphene foam supported on nickel structure (anode) and Cu foil (cathode) were immersed in 0.1 M NaOH dissolved in ethanol and subjected to 30 V for 1 h. As-synthesized dispersion initially displayed a yellow solution with basic pH ~ 8 that evolved to an orange/red color. The dispersion was repeatedly centrifuged at 10.000 rpm, filtered with 0.2 μm micro-disk filter, dried, and re-dispersed in water until reaching a purified solution without being dialyzed or chromatographed. Two control experiments were performed in order to compare to as-synthesized GQDs. Control # 1 (CN1) was performed an electrochemical cell formed by a graphite rod (anode) and Cu electrodes immersed in 0.1 M NaOH dissolved in absolute ethanol (99 % v/v) electrolyte and subjected to 30 V during 60 min. As-synthesized dispersion exhibits basic pH ~ 11. Control #2 (CN2) was obtained through an electrochemical cell formed by a graphite rod (anode) and Cu electrodes immersed in ultrapure water and subjected to 100 V during 240 min. As-synthesized dispersion exhibited neutral pH ~ 7.

**Characterization.** UV-vis absorbance spectra were recorded using a Perkin Elmer LAMBDA 1050+ UV/Vis/NIR Spectrophotometer. Fluorescence emission and excitation spectra were acquired using a Horiba Fluorolog3 and a standard quartz cuvette. UV-vis and fluorescence samples were diluted in a 1/100 dilution using absolute ethanol. Attenuated Total Reflectance-Fourier Transform-Infrared (ATR-FT-IR) spectra were obtained using a Perkin Elmer (Beaconsfield, UK) Spectrum Two instrument with a Universal ATR module. XPS data were collected using a Thermo Fisher Scientific Model K-alpha+ with Al Kα radiation at 1486.69 eV (150 W, 10 mA), charge neutralizer and a delay line detector (DLD) consisting of three multi-channel plates. Survey spectra were recorded from −5 to 1350 eV at a pass energy of 150 eV (number of sweeps: 2) using an energy step size of 1 eV and a dwell time of 100 ms. High resolution spectra for C 1s and O 1s were recorded in the appropriate regions, using an energy step size of 0.1 eV. X-ray diffraction (XRD) was acquired using a PANalytical X´Pert PRO diffractometer (40 kV, 40 mA) in the $\theta-2\theta$ Bragg-Brentano geometry. Crystalline structure characterization of Ni foam and Ni foam/G electrodes was performed by A 2θ range was selected between 10° and 70°, with steps of 0.02° and an integration time of 59 s per step, and 300 μm spot size. Scanning Electron Microscopy (SEM) on the Graphene/Nickel as well as Nickel Foams was performed by a FEI Quanta 200 at 10kV. Imaging of nanostructures and structural characterization was performed by High Resolution Transmission Microscopy (HRTEM) using an aberration (Objective Lens) corrected Titan[3] 60-300 (ThermoFisher) microscope operating at 80 kV at room temperature. Atomic Force Microscopy (AFM) images were obtained using a Nanoscope V (Veeco) microscope in tapping mode. Raman spectra of GQDs were recorded on an as-prepared SERS substrate in order to enhance the Raman signal. The SERS substrate is prepared by drop-cast deposition a solution of Au NPs protected by PVP onto the glass slide. Then, 20 μL of 1mg/ml GQDs dispersions is drop-casted in order to perform the measurement. A Horiba Jobin–Yvon Raman spectrometer LABRAM HR800 coupled to an optical microscope was used with a He-Ne laser (632.8 nm). The samples were investigated in several random zones to corroborate the homogeneity of material distribution on the support, using a 100x objective lens (NA 0.9), with 5 s acquisition and 5 s of spectra averaged, in the range of 1000 cm$^{-1}$ and 3000 cm$^{-1}$. Graphene-Nickel Foams were measured in the same optical set-up using a 532 nm laser, a 100x objective lens (NA 0.9), with 5 s acquisition and 5 s of spectra averaged, in the range of 100 cm$^{-1}$ and 3500 cm$^{-1}$.

**RESULTS AND DISCUSSION.**

**Characterization of the Electrochemical Synthesized GQDs.** Figure 1A exhibits the evolution of the UV-vis absorbance measured at different times during and post-synthesis as indicated. At the initial stage, the spectrum exhibits just a single absorbance band located at ~200-208 nm attributed to C-C groups of the solvent. During the first 60 min of synthesis, a prominent band at ~276 nm alone with a small shoulder at ~225 nm was observed. As the synthesis continues from 1 to 6 h, the peak at 276 broadens and the spectrum seems overlapped above 320 nm. The band at 276 nm has been observed in that exact position for carbon structures obtained from electrochemical trimming of GO and hydrothermal synthesis and assigned to typical π to π* transitions.[22,23] Before the 48 h after synthesis, few changes were noted including the color change of the dispersion from yellow to orange (not shown), the band at 276 nm broadens, and a new band located at 245 nm appeared. At 48 h post-synthesis the dispersion turned reddish, the entire spectra became flattened and mounted, and the band at ~331 nm broaden and red shifted toward ~366 nm. The bands observed at ~331 and ~366 nm can be ascribed to n-π* electronic transition corresponding to carbonyl groups. Figure 1B exhibits the emission spectra measured at 48 h post-synthesis of GQDs after sweeping the excitation wavelength from 205 to 390 nm as indicated. There are three distinct emission bands at ~345, ~400 and ~500 nm assigned to C=C $sp^2$ domains (first 2 bands) and functional groups (i.e.; -COOH) for the latter band. Bands located between 340-400 nm emit due to quantum confinement of $sp^2$ domains within the carbogenic core.[37] As long as the excitation wavelength increases above 360 nm, the emission intensity rises for modes associated with functional groups located at the edge or defects of carbon nanostructures.[38] There is a general agreement that emission of carbon nanostructures can be divided into two main processes described as intrinsic and extrinsic emission corresponding to the C=C bonds within the carbogenic core and functional groups located at edge/surface states, respectively.[39] Therefore the main contribution to the emission spectra corresponds to extrinsic emission associated with functional groups attached to $sp^2$ π-conjugated domains within the carbogenic core. The origin of extrinsic emission has been well-determined experimentally by deactivating (i.e.; acidic pH) oxygen-containing groups at the edge/surface of carbon structures.[39] Figure 1C shows HR-TEM images revealing mostly rounded GQDs with a prominent 7.6 ± 1.5 nm average size population as indicated by the histogram (Fig. 1D). We performed statistics and encountered, besides the ~8.0 nm, a minor population of ~3.0 nm diam. GQDs. It should be pointed out that since there are two sources of carbon, two electrochemical processes may occur simultaneously at the anode as follows: an electrooxidation of small molecules such as ethanol and electrooxidation at the vast conductive graphene surface. Therefore, we assigned these small carbon nanostructures to the electrochemical oxidation of ethanol occurring at the surface of graphene and/or defects (bare Ni) as recently determined by our group.[12] In Figure 1C, there is a zoom-in HR-TEM image of a single ~8.0 nm GQD indicating 0.21 nm inter-planar atomic distance which perfectly matches with the (100) hexagonal lattice spacing present in the graphene structure.[40]

However, more complex GQDs structures were found within similar sizes that needed further discussion (vide infra).

**Cubic Carbon Nanostructures.** Figure 2 A-C (top panel) shows HR-TEM images of the seemingly cubic structures found within the as-synthesized GQDs. The corresponding fast Fourier transform (FFT) performed on each GQD is shown in bottom panel. Figure 2 D and E illustrate a top-view layer of HCP (hexagonal closed packed) and FCC (face centered cubic) unit cells for each atom configuration; respectively. Within our product some cubic structures were also found. These types of cubic structures have different diffraction patterns which have been observed in other electrochemical and solvothermal synthesis, however, most of the time a detailed description of the product has been overlooked.[41] These types of structures may comprise NaOH, $Na_2O$, Cu, CuO, Ni, or NiO aroused from the electrosynthesis performed. However, XPS survey analysis in Figure S1 and Table S1 (Supporting Information) exhibited no traces of Cu, Ni, CuO, or NiO within the analyzed sample, prior to HR-TEM characterization. In addition, the simulated electron diffraction, for all the possible compounds mentioned above, did not match any of our experimental results as depicted in Figure S2 and Table S2. Moreover, these types of nanostructures did not appear in the control experiments (vide infra) neither in previous reported electrochemical synthesis performed by our group under nearly identical conditions on bare Ni foam electrodes. Another control experiment performed with graphite (instead of graphene) submerged in ethanol under the same voltage and time conditions did not produce any of the aforementioned nanostructures. It has been recently reported some new forms of carbon named morphed graphene (RH6-II) which exhibit characteristic cubic symmetry.[42,43] Although this type of carbon structure exhibited reflections at 3.5 Å (110), 2.32 Å (101), 2.02 Å (300), 1.93 Å (20-1), 1.75 Å (220), and 1.69 Å (211),[44] they do not match the observed diffractions in our system. Graphene can also evolve into complex forms of carbon nanostructures, during the CVD process, including different cubic diamond-like structures as recently reported by Tanigaki and co-workers.[45] So far, we ruled out the possibility of being some of those structures. Finally, our experiments and the literature lead us to conclude about the existence of few layer GQDs with possible stacking faults aroused from CVD graphene grown on Ni anodes.

Our hypothesis consists in that the observed GQDs are composed of a multilayered graphene structure piled up in an ABC…ABC stacking order. This configuration is a modification of AB…AB hexagonal structure which is defective by stacking fault. This argument is based on the fact that electron diffraction patterns relay upon the type of stacking, number of layers, and direction of the incoming beam. For monolayer graphene, carbon atoms assemble in a hexagonal lattice, adopting a different electronic configuration than isolated carbon atoms. The electronic distribution involves the promotion of one of the 2s electrons to a 2p state, leading to the formation of $sp^2$ sites in the x-y plane, with the remaining fourth electron held in a $2p_z$ state initially decoupled from the x-y plane; displaying an hexagonal simple lattice with diffraction planes separated by 0.21 nm in direct space. However, when multiple layers are stacked, the electronic environment changes as for example "bilayer graphene" may adopt two possible stacking configurations including AB (hexagonal, p63/mmc #194, a = b = 2.46 Å, c = 6.70 Å, α = 60°, β = 120°) and AA' (orthorhombic, Fmmm #69, a = 2.46 Å, b = 4.26 Å, c = 6.88 Å, α = β = γ = 90°). The stacking of two graphene layers produces an effective overlap of $2p_z$ orbitals at the α (β) sites. As a result, the $2p_z$ orbitals of the α (β) sites with no overlapping have lower energy than the overlapped ones. HR-TEM images of these nanostructures appear as a hexagonal lattice with 0.24 nm lattice parameter,[46] as well as observed in a hexagonal graphene lattice under STM measurements.[47] Trilayer graphene may adopt a rhombohedral symmetry (ABA stacked) or triangular symmetry (ABC stacked). Few layers graphene can be stacked in different configurations being the sequence AB…AB within an HCP (hexagonal closed packed) and ABC…ABC within FCC (face centered-cubic) structure.[48] In conclusion, multilayered carbon structures with seemingly cubic symmetry could be associated with this packing configuration (ABCABC…ABC) which forms an FCC lattice with the layers being normal to the direction [1,1,1]. Guerrero-Avilés et *a*l.[49] recent demonstrated that the energy difference between AB and ABC stacked graphene is less than 10 meV/nm$^2$. Remarkably, this slim margin suggests that achieving one stacking configuration over another can be influenced by introducing substrate strain. This discovery challenges previous assumptions, indicating that even small adjustments to the experimental setup can favor these less common ABC stacking configurations over the more prevalent AB stacking, resulting in an augmented prevalence of ABC-type configurations.[31,50] Substrate curvature is intricately integrated into the topology of the Ni foam utilized in the CVD synthesis, potentially explaining the elevated occurrence of ABC-stacked graphene due to stacking faults.

To further confirm the presence of few-layer graphene (FLG) within the source carbon nanostructure, we performed SEM, XRD, and Raman on the nanocarbons. Figure 3 A, B shows SEM images before and after CVD graphene grown on Ni scaffold, respectively indicating noticeable changes on the surface of the foam. Figure 3 C shows XRD diffractograms of graphene grown on Ni foam with two prominent peaks at 44.5º and 51.8º associated with the (111) and (200) planes of Ni, respectively corresponding to an FCC structure. There is also a small G peak (evidenced as an inset) at 26.6º associated with (002) plane and inter-plane separation of 0.34 nm of graphene. XRD spectra for GQDs and NaOH is shown in Figure S3 (Supporting Information) which exhibited two peaks at 26.1 º and 23.4º corresponding to 0.34 nm and 0.38 nm interlayer spacing. This spacing is in agreement with d-spacing reported for graphene and reduced graphene oxide, respectively.[51] Additional information is provided by the Raman spectrum in Figure 3D that shows the presence of D, G, G* and 2D bands at ~1340, ~1570, ~2440, ~2690 cm$^{-1}$, respectively. The intensity ratio between the G and 2D peak ($I_{2D/G}$), the full-width at half-maximum of the 2D peak (fwhm$_{2D}$), and the deconvolution of the 2D band act as finger prints for the number of layers and the overall quality of graphene. For instance, Figure 3E shows the deconvolution of 6 peaks at ~2663, ~2680, ~2691, ~2701, ~2716, ~2734 cm$^{-1}$ enclosed within the 2D band. This clearly indicates a multilayer structure of at least 2-3 graphene layers.[52] Graphene grown on Ni exhibited an $I_{2D/G}$ of 0.83 and an fwhm$_{2D}$ of 61 cm$^{-1}$ characteristic of few layer graphene (FLG).[52] Calculated $I_{D/G}$ shows a value of 0.033 being an indicator of low defects and overall crystallinity of the structure.[53] Raman spectroscopy over the obtained product is shown in Figure S4 (Supporting Information) and exhibited an $I_{D/G}$ of 2 which is consistent with 2-7 nm crystallite size. Peak deconvolution was performed into 4 peaks D*, D, D'', and G observed at 1150, 1335, 1500, and 1595 cm$^{-1}$; respectively. An overtone peak normally found at 1620 cm$^{-1}$ for D' was not observed. D peak deconvolution in different modes is associated with defects into the basal plane of the structure, commonly find in GO (graphene oxide) Raman spectra.[51,54] None of those peaks were found in pristine graphene as those signatures are associated with high degree of oxidation that is needed to rip off the graphene structure in the electrochemical oxidation.

It is expected to find FLG grown on Ni since the carbon atoms segregate and pile up at the Ni surface during the cooling down step at the end of the CVD synthesis.[55] This is also consistent with the work of Messina *et* al. who performed Raman statistics to demonstrate that graphene on Ni usually grows into FLG in a mixture of AB staking and rotated configurations.[36] Figure S5 (Supporting Information) shows AFM cross-sections GQDs heights ranging from ~0.4 to 2.1 nm and an average height of ~0.78 nm indicating the stacking of 2 to 6 layers. Interestingly, all the analyzed cross-sections revealed to be multiple of 0.34 nm confirming that the carbon precursor was effectively graphene. So far, the relationship between height and lateral dimension obtained from AFM and HR-TEM; respectively suggests a disk-shape graphene nanostructure.

**Chemical Environment of the GQDs.** Figure 4 compares deconvoluted XPS spectra between graphene coating on Ni foam before the electrosynthesis (A and B) and the resulting GQDs (C and D). Figure 4 A and B show C 1s and O 1s XPS spectra measured for graphene on Ni used as a working electrode. For C 1s, it shows a prominent band at 284.0 eV characteristic of carbon $sp^2$- hybridization. A minor band at 284.8 eV is assigned to carbon in a $sp^3$-hibridization which is also consistent with two relatively minor peaks at ~ 285.7 and ~ 288.0 eV assigned to C-O and C=O; respectively. For O 1s XPS spectra in Figure 4B, two signals at 531.3 and 532.7 assigned to C-O, and C=O groups were recorded. Unlike graphene on Ni, the C 1s for GQDs shown in Figure 4C reveals two prominent bands at 284.0 and 285.0 eV assigned to C=C bonds characteristic of $sp^2$- and $sp^3$-hybridization; respectively. There is another intense band recorded at ~287.0 eV assigned to epoxide groups (C-O-C). Additionally, there are two relatively smaller peaks at ~ 286.0 and ~ 289.0 eV associated with C-O and C=O; respectively. Figure 4 D shows deconvoluted O 1s XPS spectra dominated by COOH, C-O-C, and C=O groups evolved at 531.0, 533.0, and 534.5 eV. A minor peak seen at 536 eV is assigned to Na Auger which may correspond to byproducts of the synthesis.[51] XPS analysis confirms that graphene on Ni is comprised of $sp^2$ carbon structure mainly populated by C=C $sp^2$ bonds expect for defect areas that may have been oxidized after the CVD growth. Upon electrochemical cleavage, the graphene structure is highly oxidized and tore apart into smaller domains in the range between 3-8 nm diam. Comparing Figure 4 A and B with C and D, we can discern a notable difference in the distribution of functional groups. GQDs exhibit predominant C-O-C and COOH groups due to the electrooxidation of ethanol. These specific functional groups play a significant role in enabling the GQDs to effectively disperse in both, ethanolic and aqueous solutions. Control experiments later confirmed (vide infra) that electrochemical cavitation of graphene performed in nanopure water led to precipitation of the product indicating poor dispersion in polar solvents. Figure 4E shows ATR-IR spectra for GQDs with bands located at ~1075/846, ~1429, ~1570, 2920-2850, and ~3300 cm$^{-1}$ corresponding to C-O-H stretching / C-O-C epoxy stretching, C-H, C=C, sp$^2$/sp$^3$ C-H stretching, and -OH groups; respectively. When compared with carbon quantum dots (CQDs) obtained from a bare Ni foam (i.e., no graphene coating), we observed an inversion in the intensity of the bands at 1570 and 1427 cm$^{-1}$, assigned to C=C and -C-O, for both carbon nanostructures. This is consistent with Wang *et al.* who noted that as graphene oxide (GO) becomes reduced, all the oxygen-containing band intensities diminish, except for the former band at 1570 cm$^{-1}$ associated with the restoration of the π-system (C=C vibrations) in graphene.[56] The same group further confirmed their results with solid-state $^{13}$C NMR, which exhibited the predominance of graphitic sp$^2$ domains upon fully reducing the GO.

**Control Experiment # 1 (Carbon Nanostructures 1, CN1): The Role of the Working Electrode.** In this control, the 3D graphene foam was replaced by a graphitic rod in order to study the role of the working electrode. Figure 5 corresponds to HR-TEM images for the control experiment #1 named carbon nanostructures (CN1) performed using a graphite rod instead, but keeping the other parameters the same. Figure 5 A and B show an HR-TEM image with low magnification displaying an uneven covered grid along with agglomerated carbon structures. Figure 5C shows clouds- or islands-like features with dendritic fringes. Between these nano-islands there is a large amount of different carbon structures for analysis. Figure 5C displays the corresponding FFT from Figure 5B showing the hexagonal symmetry and crystallinity of the observed graphitic structure. Figure 5D display an HR-TEM image of the area between nano-islands with a high concentration of crystalline carbon nanoparticles, their corresponding size distribution is displayed as an inset. Figure 5E and F show the corresponding ROI marked with red and blue respectively. Lattice fringes of 0.21 nm corresponding to the (10-10) of graphite are systematically observed. Table S3 (Supporting Information) shows the observed diffraction spots, their lattice spacing, and the angles between them. Results indicate the presence of relatively large and agglomerated graphitic-type structures that may seem to be comprised of various different graphitic carbon structures. The counterproductive use of graphitic rods instead of CVD graphene is evident in terms of the overall quality and homogeneity of the outcome product. So far, using graphene foam as a precursor ensures crystallinity and overall control over the quality of the product with no need of dialysis or chromatography techniques.

**Control Experiment # 2 (Carbon Nanostructures 2, CN2): The Role of the Electrolyte.** This experiment was carried out in order to assess the role of EtOH and NaOH in the electrochemical synthesis. The first experiment consisted in the electrosynthesis performed in graphene-coated Ni foam but, this time, submerged in ultrapure water without NaOH. Since there is no proper electrolyte, the voltage applied needed to increase from 30 to 100 V (during 240 min) in order to achieve a considerable concentration of product for further analysis. We observed a black precipitate likely corresponding to hydrophobic graphene material exerted from the anode. Then, we replaced the anode with a graphite rod and surprisingly observed a brown color dispersion later used for further analysis. Figure 6A shows a typical HR-TEM image taken from the brown dispersion which exhibits a characteristic crosslinked features suggesting a multilayered structure. Figure 6B shows FFT performed on the spot marked in red indicating highly crystalline reciprocal space. Figure 6C shows the IFFT of the filtered reciprocal space for this crystal. Table S4 (Supporting Information) shows the indexed diffraction spots, their lattice spacing, and the angles between them for CN2. Interestingly, we found cubic like carbon structures that do not exactly match a cubic diamond structure as discussed before. Diffractions spots at 0.19 and 0.29 nm were also found in GQDs and CN1 confirming their graphitic origin. The most captivating diffraction peak is associated with the 0.25 nm spot, positioned at an angle of 88.5 degrees relative to the 0.29 nm diffraction. This arrangement showcases nearly cubic characteristics. This diffraction pattern has also been observed in the current as-synthesized GQDs supporting the hypothesis that cubic lattice features can be associated with multilayered carbon systems. Since the only source of carbon comes from the graphitic rod, we conclude that these cubic structures most likely emerge from the electrochemical cleavage of the anode subjected to high potentials. Figure S6 shows amorphous carbon nanoparticles as well as multi-walled carbon nanotubes found within this sample as part of the overall outcome product. We believe that these kinds of nanostructures emerged due to the extremely large potentials used (100 V).

**Proposed Mechanisms for the Electrochemical Synthesis.** Figure 7 shows a scheme of all the products obtained from the different experimental set-up and conditions used in this study as well as results obtained from a previous one.[12] A cartoon representing the diverse products obtained is indeed shown. The upper panel depicts the current work where the obtained product is generated by electrooxidation of ethanol and electrochemical exfoliation of CVD graphene foam generating 3.0 to 8.0 nm FLG quantum dots (GQDs). The bottom panel illustrates nanocarbons synthetized by electrooxidation of ethanol in alkaline media using a bare Ni foam as electrode, producing ultra-small 3.0 nm crystalline carbon quantum dots.[12] The left panel depicts control # 1 (CN1) in which a graphitic rod (in NaOH/EtOH) is used instead of CVD graphene. The obtained product is generated by dual electrooxidation of ethanol and electrochemical exfoliation of graphite rods producing polydisperse, relatively large, and multilayered carbon structures. The right panel shows CN2 where the obtained product is generated by electrochemical exfoliation of graphite rods in ultrapure water leading to amorphous carbon dots, multi-layered carbon structures, and multi-walled carbon nanotubes. Table 1 shows the outcome of different carbon nanostructures produced including GQDs, carbon quantum dots (CQDs), CN1 and CN2. We investigated the influence of the working electrode, electrolyte, and electrochemical conditions on the production of different types of carbon nanostructures. Since the electrochemical synthesis is performed in the presence of two distinct carbon sources (ethanol and graphene) it is reasonable to expect bottom-up and top-down mechanisms occurring simultaneously either in the production of GQDs and the control experiments (CN1 and CN2). The top-down approach, in alkaline media, takes place via the oxidation of C=C bonds usually at defect sites within the basal plane of graphene. Figure S7 A and B (Supporting Information) shows a scheme of the three electrode set-up used in this work along with cyclic voltammograms curves run with and without ethanol and graphene at the anode as indicated. Section S9 (Supporting Information) develops a qualitative analysis indicating that graphene is enhancing the electrooxidation of ethanol and partially protecting the Ni foam (from oxidation) as evidenced by the observed higher currents and overpotentials; respectively. When graphene is at the anode, this mechanism is most likely exacerbated by the generated $O_2$, which leads to hydroxyl ions to intercalate between graphene layers in a cavitation-like process along random directions in the graphene lattice.[57] Several theoretical studies indicate that epoxides are responsible for ripping graphene layers off converting epoxide into carbonyl groups and effectively breaking C=C bonds in the process at the defect sites of graphene.[35] Therefore, the proposed mechanism for the electrochemical cleavage of graphene depends upon two main processes occurring simultaneously at the anode. This includes the electrooxidation of ethanol[58] and exfoliation-like process preferentially.[35]

**Table 1.** The Table indicates all the possible carbon structures obtained after alternating the structure and composition of the anode as well as the electrolyte solution.

| Carbon structures | Anode | Cathode | Electrolyte | VDC (V) | Time (min) |
|---|---|---|---|---|---|
| GQDs | Graphene /Ni Foam | Cu | EtOH + NaOH (0.1 M) | 30 | 60 |
| CQDs | Ni Foam | Cu | EtOH + NaOH (0.1 M) | 30 | 60 |
| Control # 1 (CN1) | Graphite Rod | Cu | EtOH + NaOH (0.1 M) | 30 | 60 |
| Control # 2 (CN2) | Graphite Rod | Cu | ultrapure Water | 100 | 240 |

**CONCLUSIONS**

Our work demonstrated that the electrochemical synthesis from CVD graphene led to subdomains called GQDs whose crystallinity and structure (few layered graphene) were preserved. A comprehensive structural characterization involving HR-TEM and XRD revealed GQDs consisting of multilayered AB…AB configurations showing a hexagonal structure as well as ABC...ABC stacking fault defects becoming a transition HCP → FCC structure. The XPS, ATR-IR, and photoluminescence studies further confirmed that GQDs comprise $sp^2$ graphene-like domains surrounded by $sp^3$ oxygen-containing groups. Although dissimilar mechanisms are activated, electrochemical exfoliation (C-C bond oxidation-breakdown) is sufficiently favored to result in a major population of ~8 nm diam. GQDs obtained from CVD graphene onto nickel foam. Control experiments employing graphitic rods yielded agglomerated and polydisperse carbon structures indicating the importance of the carbon source used in the electrosynthesis. This research demonstrated that the utilization of graphene foam as a pristine starting material ensures control over the size, structure, and overall quality of the GQDs product. Remarkably, these achievements were possible via a simple and well-known electrochemical process underlining the power of electrooxidation to retain the valuable atomic configurations of CVD graphene frameworks. More generally, we envisioned that performing electrochemistry of graphene should even enable the emergence of many exciting new heteroatom 2D quantum dots from heterostructures of van der Waals materials.

**ASSOCIATED CONTENT.**

**Supporting Information.** Survey XPS spectra and C/O content for GQDs, electron diffraction simulation for $NaO_2$, Cu, CuO, Ni, and NiO, XRD for GQDs and NaOH, Raman spectra for GQDs, additional AFM images and profiles of single and bilayer GQDs, table S6: Showing the observed diffraction spots, their lattice spacing, and the angles between them for CN1, table S7: Showing the observed diffraction spots, their lattice spacing, and the angles between them for CN2, amorphous carbon nanoparticles and multi-walled carbon nanotube found in Control #2 (CN2), brief description of ethanol electro-oxidation in alkaline media with Graphene/Nickel foams as working electrode.

**AUTHOR CONTRIBUTIONS**

The manuscript was written through contributions from all authors. All authors have given approval of the final version of the manuscript.

**CONFLICT OF INTEREST**

There are no conflicts to declare

**ACKNOLEWDGMENTS**

We gratefully acknowledge financial support from PICT (2019-2188), PIP-2022-2024 0001 (CONICET), and Proyecto de Incentivos 2019-X-887 from UNLP. We also acknowledge the financial support of (Grant No. 872631) and ULTIMATE-I (Grant No. 101007825). Authors would like to acknowledge the access of equipment of "Servicio General de Apoyo a la Investigación (SAI), Universidad de Zaragoza". FJI acknowledge Dr. Mauricio Llaver for the discussion on FT-IR spectrum of GQDs. SB and FF acknowledge CONICET for their doctoral scholarships.

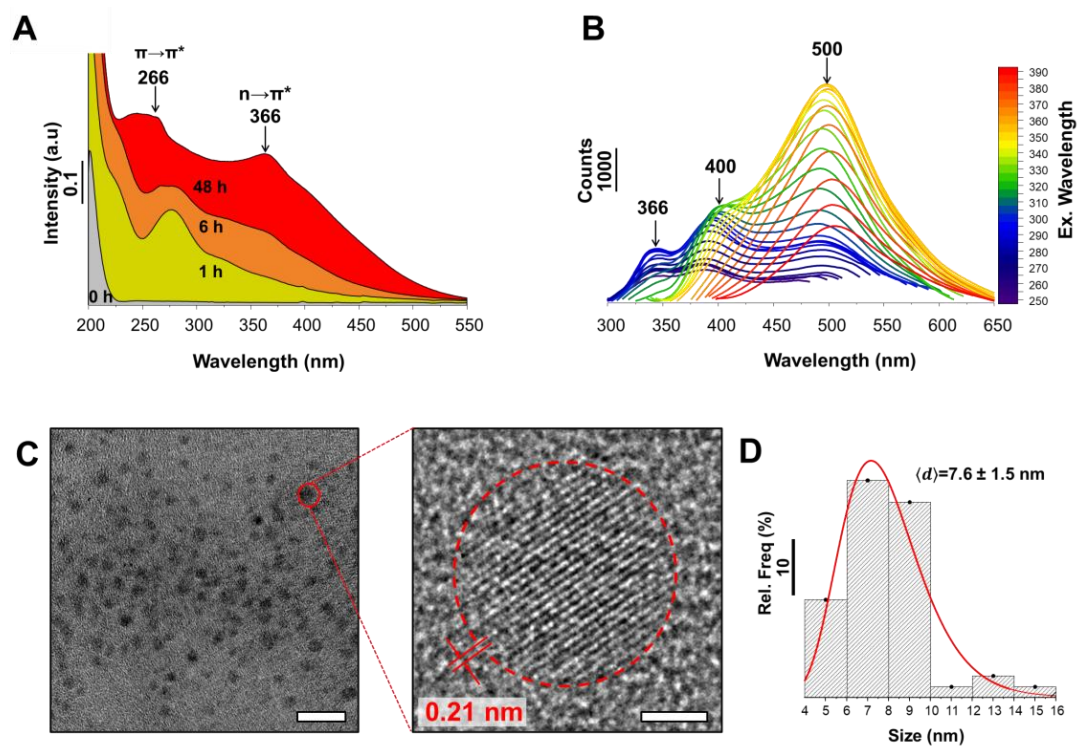

**Figure 1.** UV-vis spectra show time evolution during and after the electrosynthesis measured at the indicated times (A). Fluorescence emission spectra for GQDs measured 48 h post-synthesis (B). Survey HR-TEM image over a population of GQDs (C) along with a zoom-in showing HR-TEM image of a single GQD (D). Histogram indicating a size population over 5 nm. Scale bar in figure C and D represents 10 and 2 nm, respectively.

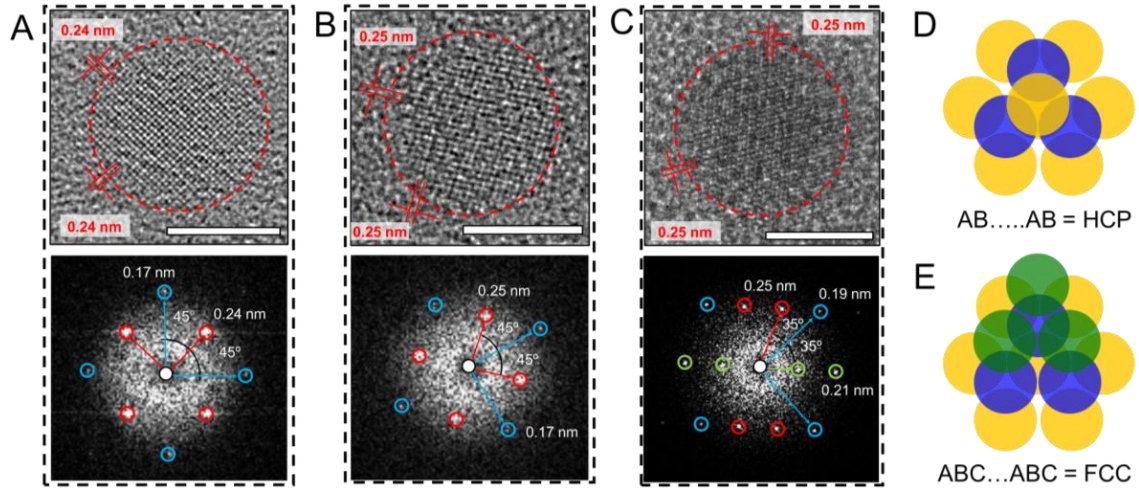

**Figure 2.** HR-TEM images (top) and FFT (bottom) of Few Layer GQDs (A, B, C). Top view of HCP and FCC stacking respectively (C, F). Yellow, blue and green spheres represent atoms in A, B, C type of layers respectively. Scale bar represents 5 nm.

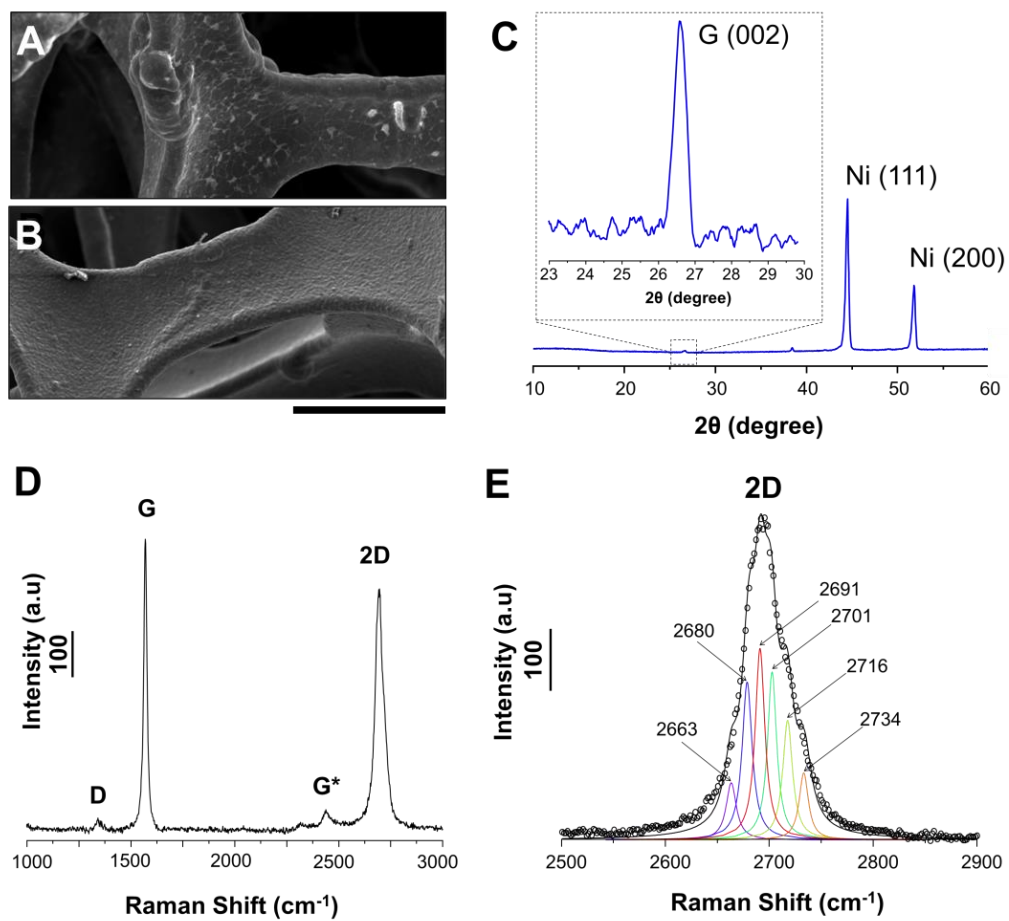

**Figure 3.** SEM Images of graphene on Ni (A) and bare Ni (B) foams, scale bar represents 100 μm. XRD diffraction pattern of graphene grown on Ni foam (C) inset shows a magnification over the 002 peak of graphene. Raman spectrum (D) and the 2D peak deconvolution (E) obtained from the working electrode (graphene grown on Ni).

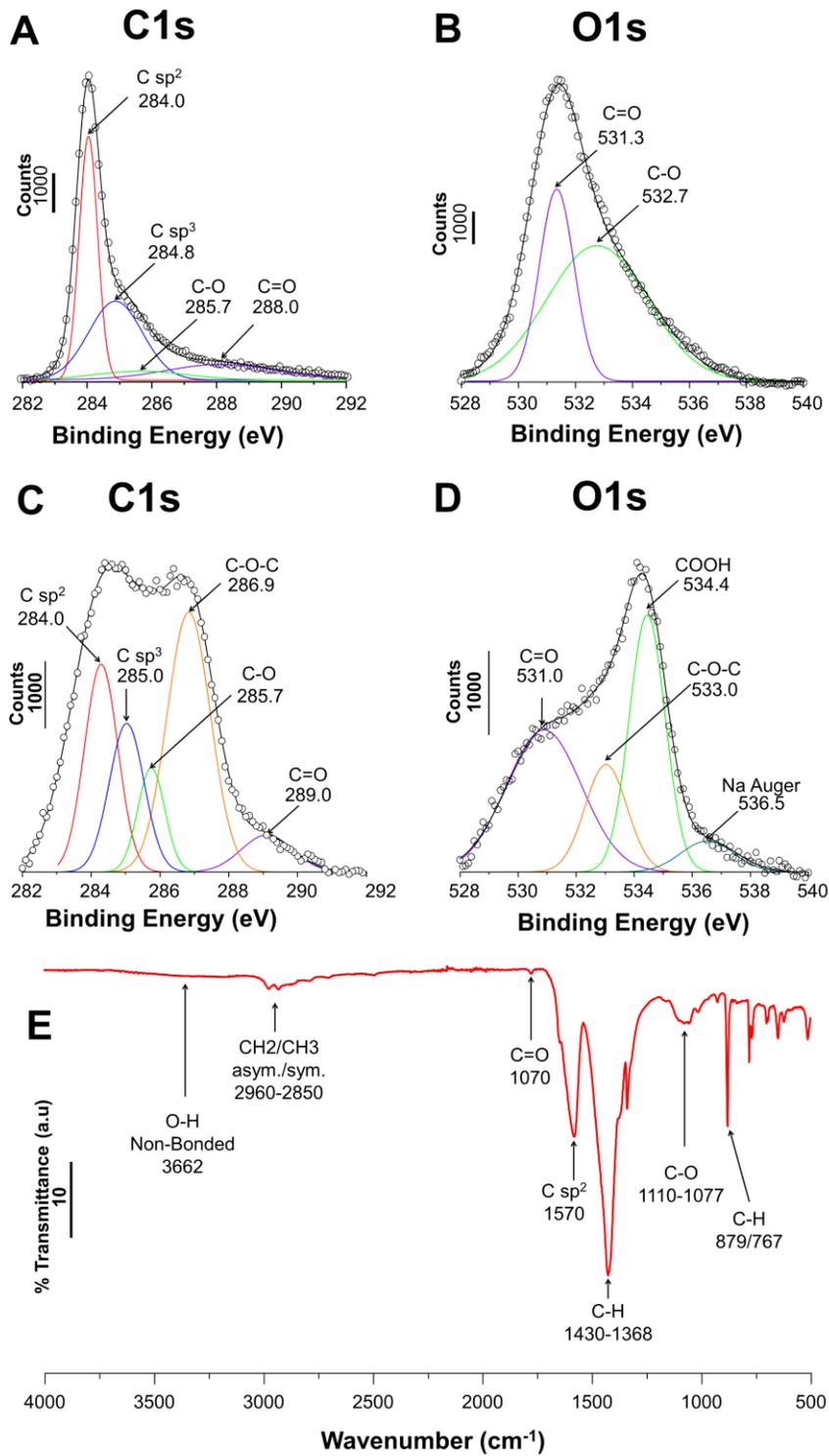

**Figure 4.** Deconvoluted C 1s XPS spectra for graphene on Ni (A). Deconvoluted O 1s XPS spectra for graphene on Ni (B). Deconvoluted C 1s XPS spectra for GQDs (C). Deconvoluted O 1s XPS spectra for GQDs (D). ATR-FTIR fro as-synthesized GQDs (E).

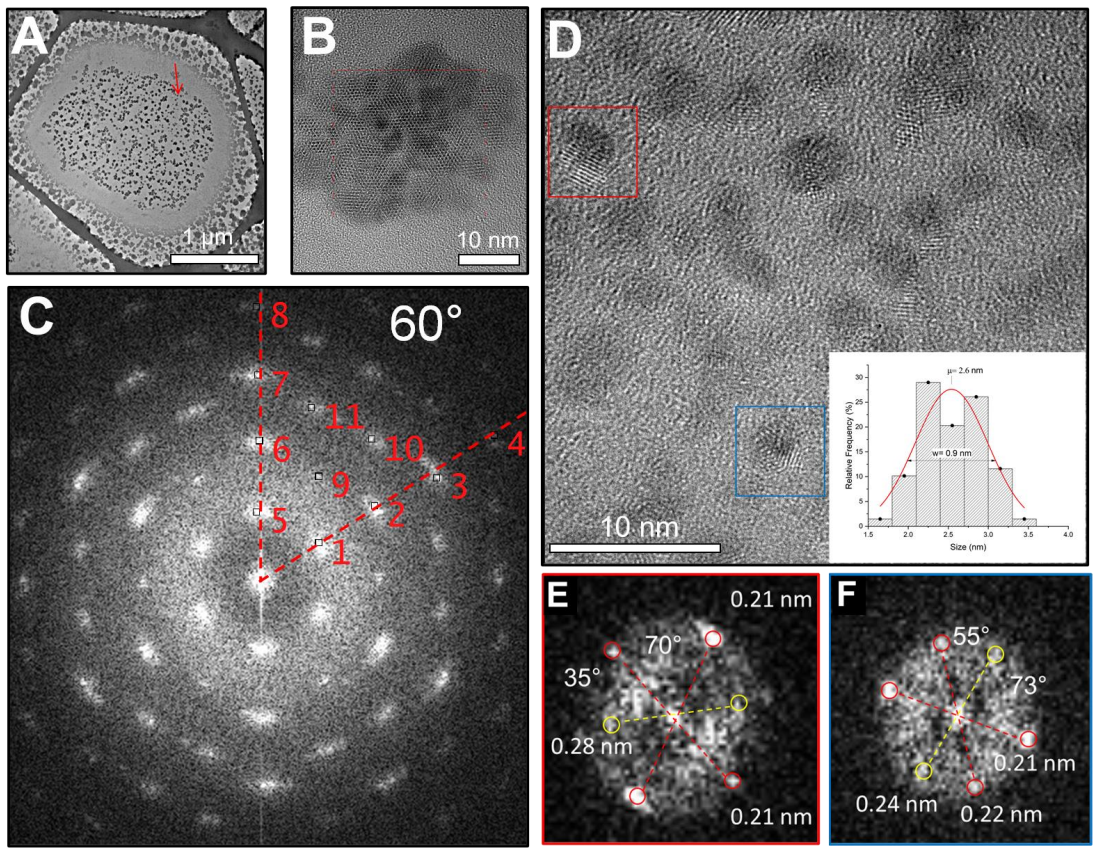

**Figure 5.** HR-TEM survey image of CN#1 showing grid coverage (A). Agglomeration of nanoparticles in small islands (nano-islands) (B). FFT of observed nano-islands in B (C). Zoom-out HR-TEM image along with a histogram of size distribution (D). FFT of the corresponding ROI marked with red (E) and blue (F) blue boxes.

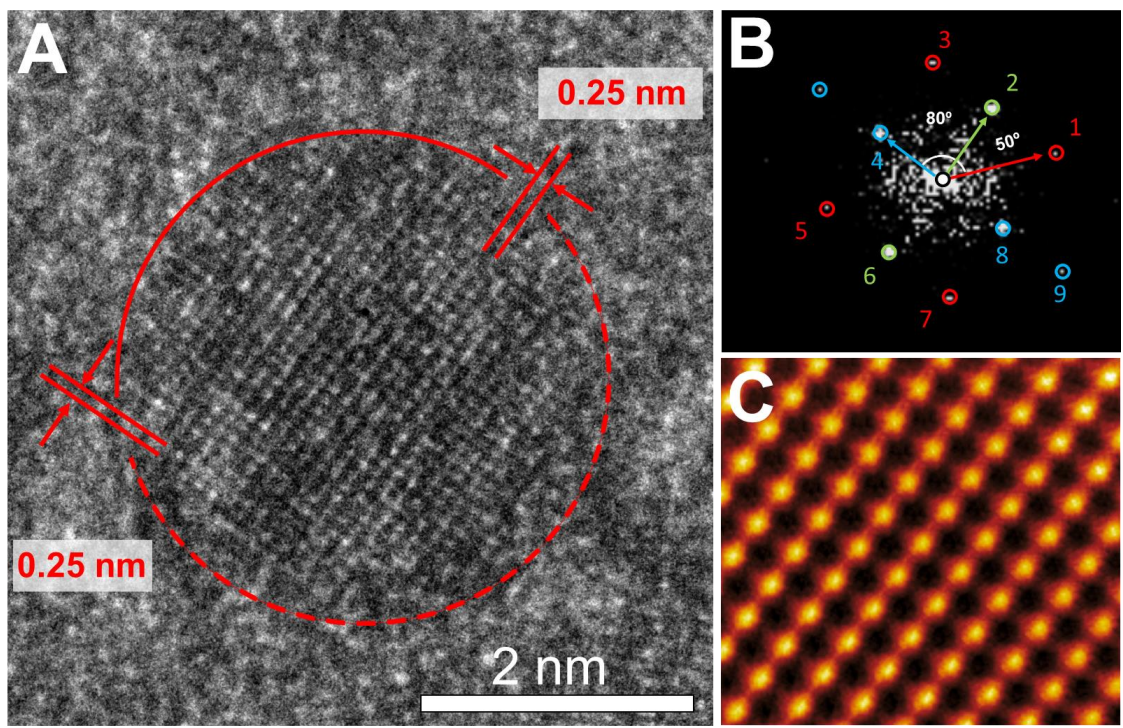

**Figure 6.** HR-TEM image of CN#2 (A) with the corresponding FFT (B). Inverse FFT of the filtered reciprocal space is shown in C.

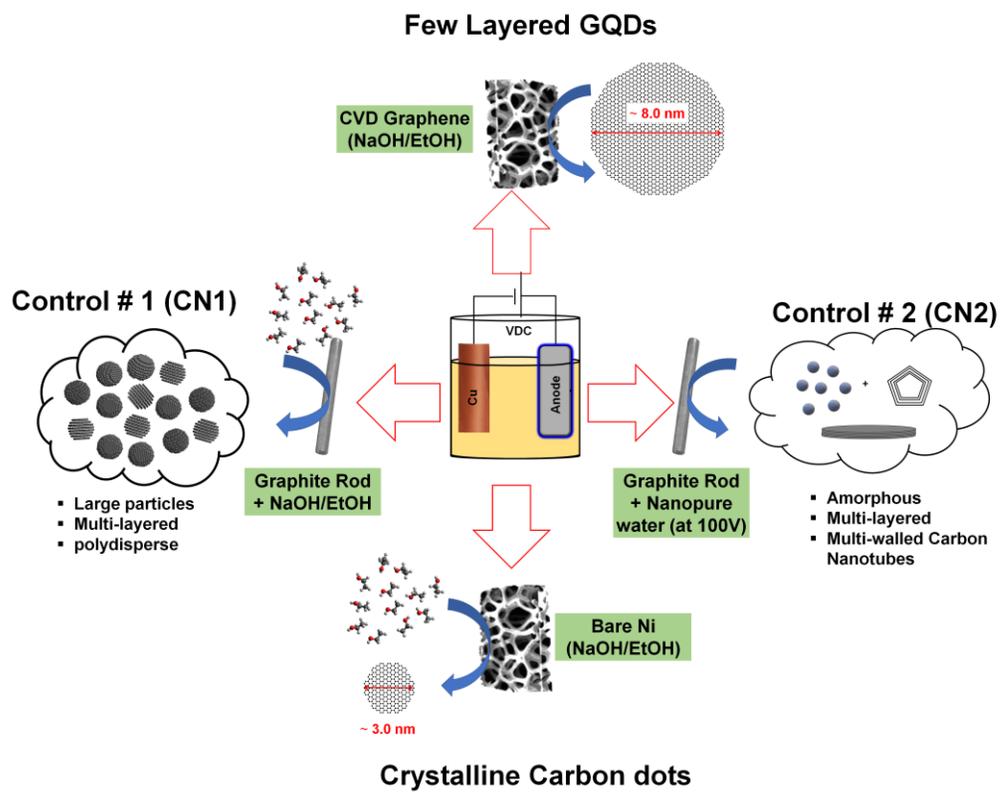

**Figure 7.** The Scheme shows the different products obtained from modifications of the electrodes and general conditions explored in this work along with a previous one from our group.